\documentclass[fleqn,twoside]{article}
\usepackage{espcrc2}

\usepackage{graphicx}

\newcommand{\be}{\begin{equation}}
\newcommand{\ee}{\end{equation}}
\newcommand{\beq}{\begin{eqnarray}}
\newcommand{\eeq}{\end{eqnarray}}

\newcommand{\AmS}{{\protect\the\textfont2
  A\kern-.1667em\lower.5ex\hbox{M}\kern-.125emS}}

\title{The ground state of three quarks
\thanks{Talk presented by Ph.\ de Forcrand.}}

\author{C.~Alexandrou\address{Department of Physics, University of Cyprus,
CY-1678 Nicosia, Cyprus},
Ph.\ de Forcrand\address[ETH]{Institute for Theoretical Physics,
    ETH Z\"{u}rich, CH-8093 Z\"{u}rich, Switzerland}
\address{CERN, Theory Division, CH-1211 Geneva 23, Switzerland} 
and O.~Jahn\addressmark[ETH]}
       
\begin{document}

\begin{abstract}
We measure the static three-quark potential in $SU(3)$ lattice gauge theory
with improved accuracy, by using all available technical refinements,
including L\"uscher-Weisz exponential variance reduction. 
Together with insight gained from 3-state Potts model
simulations, our results allow us to sort out the merits of the
$\Delta$- and $Y$-ans\"atze.
\vspace{0.8pc}
\end{abstract}

\maketitle

\section{Introduction}

The static $q \bar{q}$ potential has been studied extensively on the lattice.
In the quenched theory, the elementary ansatz 
$V_{q\bar{q}}(r) = V_0 - \frac{\alpha}{r} + \sigma_{q\bar{q}} r$
turns out to be remarkably accurate.
The next simplest system is one of 3 static quarks, from which one can gain
phenomenological insight about the forces inside a baryon. Moreover, 
the 3 quarks must be connected by 3 glue strings to form a gauge-covariant
object. These strings meet at a ``gluon junction'', which has been conjectured 
to be a non-perturbative excitation of the QCD vacuum~\cite{junction},
and might play an important role at the hadronization stage in heavy-ion
collisions.

\begin{figure}[tb]
\vspace{-0.8cm}
\centerline{\mbox{\includegraphics[scale=0.5]{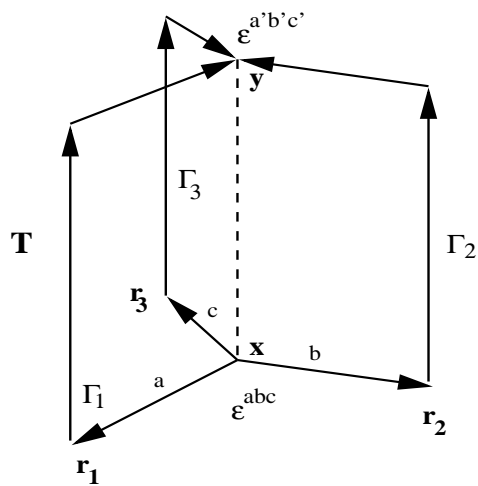}}}
\vspace{-0.8cm}
\caption{Baryonic Wilson loop with junctions at $x$ and $y$:
$W \equiv \frac{1}{6} \epsilon_{abc} \epsilon_{a'b'c'} 
\Gamma_1^{aa'} \Gamma_2^{bb'} \Gamma_3^{cc'}$.}
\label{BWL}
\vspace{-0.5cm}
\end{figure}

We extract the potential between 3 static quarks in $SU(3)$ gauge theory
from the exponential decay with $T$ of the expectation value of the baryonic
Wilson loop Fig.~1.
We aim at improving the accuracy of our earlier work~\cite{PRD} 
through several technical refinements. This new accuracy, augmented by 
insight from Potts model simulations, allows us to reach conclusions about
the merits of the $\Delta$-ansatz~\cite{Delta} and the $Y$-ansatz~\cite{Y}.

At short distances, the potential is described by perturbation theory,
which gives it the form (constant + Coulomb). 
$(i)$ The constant term is caused by UV divergences arising from the 
perimeter of the loop. For large $T$, the baryonic loop perimeter is 
$\approx \frac{3}{2}$ that of a rectangular, mesonic loop, so that the constant term in
the potential is multiplied by $\frac{3}{2}$ (for the same lattice spacing $a$).
$(ii)$ The exchange of one gluon between two of the quarks gives $\frac{1}{2}$
the one-gluon exchange term between a quark and antiquark at the same 
locations. 
Together, $(i)$ and $(ii)$ imply
\be
V_{qqq}(\vec{r}_1,\vec{r}_2,\vec{r}_3) = \frac{1}{2}
\sum_{i<j} V_{q\bar{q}}(\vec{r}_i,\vec{r}_j)
\equiv V_{qqq}^\Delta(\vec{r}_1,\vec{r}_2,\vec{r}_3)
\label{Delta}
\ee
At large distances, the $\Delta$-ansatz predicts that the potential grows
linearly with the perimeter $L_\Delta$ of the quark triangle:
$V_{qqq} \propto \sigma_{q\bar{q}} \frac{L_\Delta}{2}$,
so that Eq.(\ref{Delta}) still holds. It is derived from a model of 
confinement by center vortices using a beautiful topological argument~\cite{Delta}.
The $Y$-ansatz predicts instead $V_{qqq} \propto \sigma_{q\bar{q}} L_Y$,
where $L_Y$ is the minimal length of the 3 flux tubes necessary to join the
3 quarks at the so-called Steiner point.
It is derived from strong coupling arguments~\cite{Y}, and is consistent with
the dual superconductivity confinement scenario~\cite{Abelian}. Since $L_Y > \frac{L_\Delta}{2}$
for all 3-quark geometries, the $Y$-ansatz predicts a steeper potential
\be
V_{qqq}^Y(\vec{r}_1,\vec{r}_2,\vec{r}_3) = 
V_{qqq}^\Delta(\vec{r}_1,\vec{r}_2,\vec{r}_3)
+ \sigma_{q\bar{q}} (L_Y - \frac{L_\Delta}{2})
\label{Y}
\ee
with $V_{qqq}^\Delta$ as per Eq.(\ref{Delta}).
Both ans\"atze are constrained to reproduce the diquark limit
$\vec{r}_j \rightarrow \vec{r}_k$, $V_{qqq}(\vec{r}_i,\vec{r}_j,\vec{r}_k) 
\rightarrow V_{q\bar{q}}(\vec{r}_i,\vec{r}_j)$
exactly, and therefore contain no free parameter once $V_{q\bar{q}}$ is given.
In this respect we differ from the analogous lattice study of \cite{Japan},
where $\sigma_{q\bar{q}}$ and $\sigma_{qqq}$ are fitted separately and
therefore not strictly equal.

\section{Technical refinements}

Because the difference between the $\Delta$- and the $Y$-ans\"atze is very
small ($1 \leq \frac{L_Y}{L_\Delta / 2} \leq \frac{2}{\sqrt{3}}$), high accuracy in the 
determination of $V_{qqq}$ is mandatory. 
The main difficulty at large quark separation is the contribution of excited
$qqq$ states. Besides smearing the spatial links as in \cite{Japan}, we use 
three additional techniques to control these systematic errors. 
$(i)$ We form a variational basis with different
junction locations ($x, y$ in Fig.~1). $(ii)$ We use multihit for the timelike
links. $(iii)$ We generalize the multilevel algorithm of 
\cite{LW}, originally proposed for Polyakov loop correlators, to baryonic
Wilson loops. This method provides a variance reduction exponential in $T$,
which allows us to extract the potential from longer loops, with crucially
improved filtering of excited states.

\section{Results}

A sample of current results based on 160 analyzed $16^3 \times 32$ configurations at 
$\beta=5.8$ and $6.0$ is shown in Fig.~2 (3 quarks in an
equilateral triangle). They are compatible with our earlier measurements~\cite{PRD},
but the reduced errors now clearly show that neither ansatz gives a proper
description of the potential. It approaches the $\Delta$-ansatz at short
distances as expected, but seems to rise faster, perhaps as fast as the
$Y$-ansatz, at large distances. Furthermore, the larger the quark separation,
the more our variational groundstate favors junctions located near the
Steiner point.

\begin{figure}[tb]
\centerline{\mbox{\includegraphics[height=9cm]{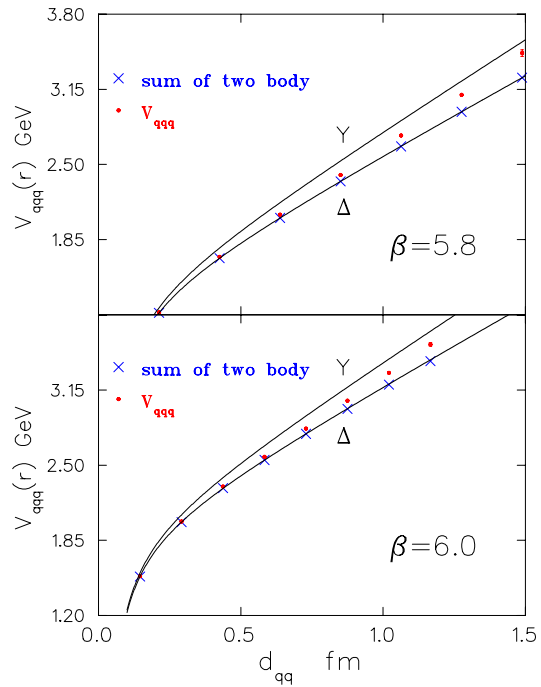}}}
\vspace{-0.8cm}
\caption{Static potential $V_{qqq}$ vs quark separation at $\beta=5.8$ and 6.0.
Also shown are the $\Delta$- and $Y$-predictions Eqs.(\ref{Delta}) and (\ref{Y}).}
\label{Vqqq}
\vspace{-0.6cm}
\end{figure}

To elucidate the asymptotics of the potential, we turned to the 3-state
Potts model. This toy model preserves the center degrees of freedom of $SU(3)$
and is thus more likely to agree with center-vortex-based predictions of the
$\Delta$-ansatz. In this model, we measured the 3-spin correlation,
after adjusting the coupling to match the $\beta_{SU(3)}=6.0$ correlation
length.
High-precision cluster Monte Carlo results were obtained for multiple
3-spin geometries, in $2d$ and $3d$.
In all cases, the 3-spin correlation behaved just like in $SU(3)$, falling
``in-between'' the $\Delta$- and the $Y$-ans\"atze. But we could establish
that the potential was rising asymptotically $\propto L_Y$.
Large separations are required to see this. An example is shown
in Fig.~3, where the change in action density caused by the 3 sources $(a)$
is compared with the superposition $(c)$ of $q\bar{q}$ action densities $(b)$,
as predicted by the $\Delta$-ansatz. A $Y$ string pattern is visible, but the
strings are ``fat'' ($\sim 0.8$ fm) compared to their length, even though
the ``quark'' separation is $\sim 2.8$ fm.

\begin{figure}[tb]
\tabskip=0pt\halign to\hsize{\hfil#\hfil\tabskip=0pt plus1em&
\hfil#\hfil&\hfil#\hfil\tabskip=0pt\cr
\mbox{\includegraphics[height=2.6cm,trim=72 48 5 48,clip]{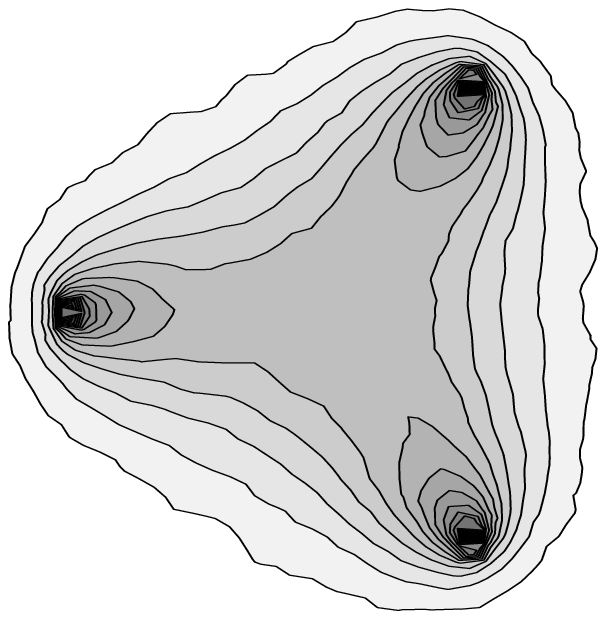}}&
\mbox{\includegraphics[height=2.6cm,trim=158 48 0 48,clip]{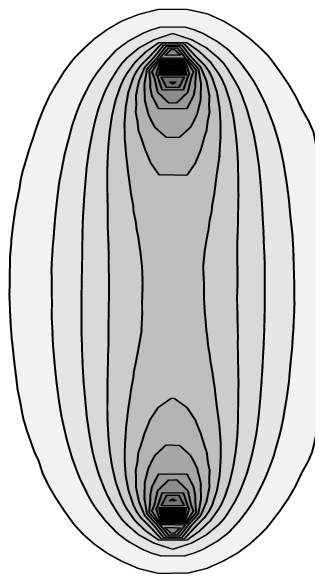}}&
\mbox{\includegraphics[height=2.6cm,trim=65 48 0 48,clip]{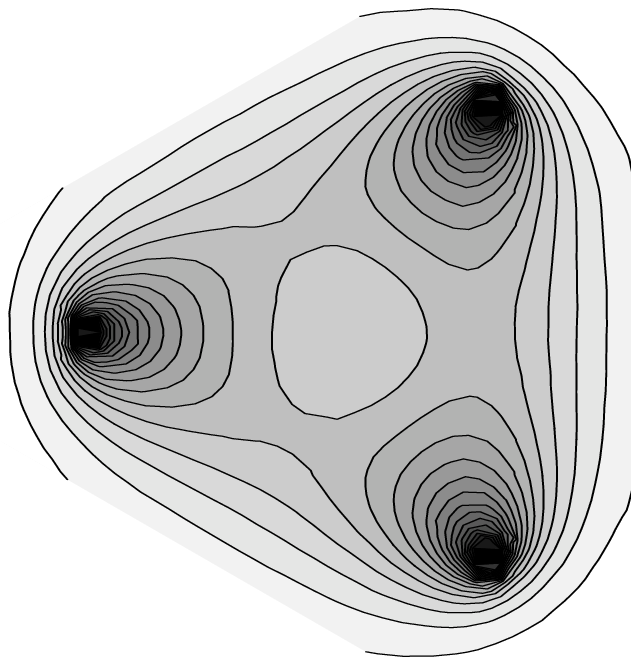}}\cr
(a)&(b)&(c)\cr}
\vspace{-0.6cm}
\caption{Action densities in the $2d$ Potts model: (a) $qqq$, (b)
  $q\bar q$; (c) $\Delta$ prediction, 
i.e. superposition of 3 $q\bar q$ densities.
The $qq$ distance is $\sim 2.8$~fm. }
\label{Potts_density}
\vspace{-0.7cm}
\end{figure}

Remarkably, the approach to $Y$-asymptotia is well described by the same
ansatz in the $2d$-, $3d$- Potts model and in $SU(3)$. As shown in Fig.~4,
the measured value of the 3-spin correlation (or $V_{qqq}$) falls short of 
the $Y$-prediction by an amount
\be
V_{qqq} - V_{qqq}^Y \approx c_0 + c_1 \exp(-L_\Delta / c_2)
\ee
where $c_0, c_1, c_2$ depend on the $qqq$ triangle geometry.
For the $SU(3)$ equilateral case, $c_2 \sim 2$~fm.

\begin{figure}[tb]
\centerline{\mbox{\includegraphics[width=6.5cm,height=3.7cm]{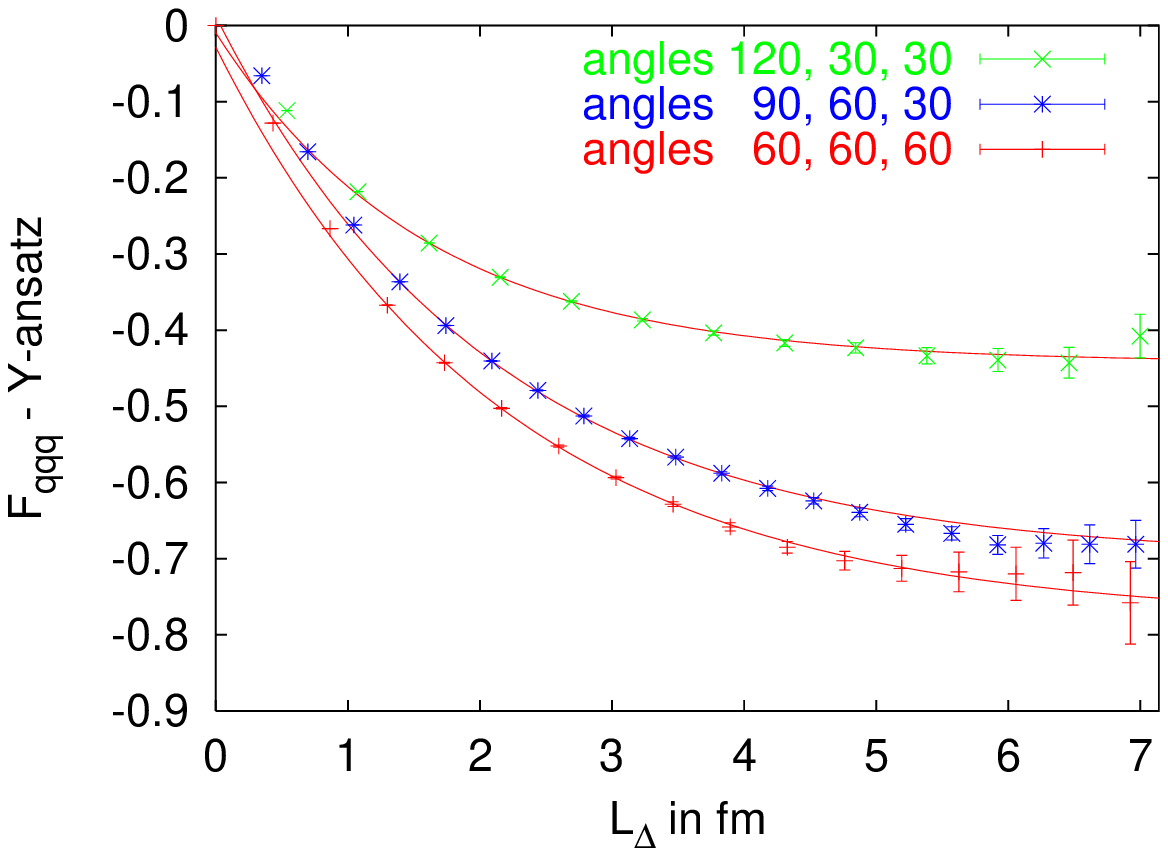}}}
\centerline{\mbox{\includegraphics[width=6.5cm,height=3.7cm]{Y-Vqqq_c.ps26}}}
\vspace{-0.7cm}
\caption{Amount by which $V_{qqq}$ misses the $Y$-prediction vs perimeter
of the quark triangle, for the $3d$ Potts (top) and $SU(3)$ (bottom) cases.}
\vspace{-0.7cm}
\end{figure}

\section{Conclusions}

Our results show that the baryonic static potential is neither
of the $\Delta$- nor of the $Y$-type. It approaches the $\Delta$-ansatz
at short distances, but rises like the $Y$-ansatz at large distances.
For an equilateral $qqq$ arrangement, departure from the $\Delta$-ansatz
is not significant until $d_{qq} \sim 0.7$ fm, so that the $\Delta$-ansatz
may be the more relevant one for quarks confined inside a hadron.

The delay in the onset of the $Y$-behavior is presumably caused by
fluctuations in the location of the junction. In the Potts model,
locations $x_J$ away from the Steiner point are suppressed only if the
associated length $L(x_J)$ of the ``glue'' strings exceeds the minimum $L_Y$ by
an amount comparable to the correlation length $\xi$. For a junction
coinciding with one of the ``quarks'', for example, the inequality
$L(x_J) - L_Y \geq \xi$ 
becomes satisfied only if $d_{qq} \geq 1.7$ fm.

Unfortunately, the $Y$-behavior at large distances does not help to rule out
the center vortex picture of confinement.
Rather, the failure of the $\Delta$-type prediction of \cite{Delta} 
appears to be due to additional hidden assumptions about the independence of
certain linking numbers.


\begin{thebibliography}{99}
\bibitem{junction} 
D.~Kharzeev,
Phys.\ Lett.\ B {\bf 378}, 238 (1996).

\bibitem{PRD}
C.~Alexandrou, P.~De Forcrand and A.~Tsapalis,
Phys.\ Rev.\ D {\bf 65}, 054503 (2002);
Nucl.\ Phys.\ Proc.\ Suppl.\  {\bf 106}, 403 (2002);
Nucl.\ Phys.\ Proc.\ Suppl.\  {\bf 109}, 153 (2002).

\bibitem{Delta}
J.~M.~Cornwall,
Phys.\ Rev.\ D {\bf 54}, 6527 (1996).

\bibitem{Y}
N.~Isgur and J.~Paton,
Phys.\ Rev.\ D {\bf 31}, 2910 (1985).

\bibitem{Abelian}
M.~N.~Chernodub and D.~A.~Komarov,
JETP Lett.\  {\bf 68}, 117 (1998);
Y.~Koma, E.~M.~Ilgenfritz, T.~Suzuki and H.~Toki,
Phys.\ Rev.\ D {\bf 64}, 014015 (2001).

\bibitem{Japan}
T.~T.~Takahashi, H.~Matsufuru, Y.~Nemoto and H.~Suganuma,
Phys.\ Rev.\ Lett.\  {\bf 86}, 18 (2001);
Phys.\ Rev.\ D {\bf 65}, 114509 (2002).

\bibitem{LW}
M.~L\"uscher and P.~Weisz,
JHEP {\bf 0109}, 010 (2001).

\end{thebibliography}
\end{document}